\documentstyle[graphicx,twocolumn,aps,prl,floats]{revtex}

\begin{document}

\draft

\title{Observation of the Quantum Zeno and Anti-Zeno 
effects in an unstable system}
\author{M. C. Fischer, B. Guti\'errez-Medina, and M. G. Raizen}
\address{Department of Physics, The University of Texas at Austin,
Austin, Texas 78712-1081}
\date{\today}
\maketitle

\begin{abstract}
We report the first observation of the Quantum Zeno and Anti-Zeno 
effects in an unstable system. Cold sodium atoms are trapped in a far-detuned
standing wave of light that is accelerated for a controlled duration. For a large
acceleration the atoms can escape the trapping potential via tunneling.
Initially the number of trapped atoms shows strong non-exponential 
decay features, evolving into the characteristic exponential decay behavior.
We repeatedly measure the number of atoms remaining trapped 
during the initial period of non-exponential decay.
Depending on the frequency of measurements we observe a decay
that is suppressed or enhanced as compared to the unperturbed system.
\end{abstract}


\narrowtext

Unstable quantum systems are predicted to exhibit a short time deviation from the 
exponential decay law~\cite{winter61,fonda78,niu98}.
This universal phenomenon led to the prediction that 
frequent measurements during this non-exponential period could inhibit decay of 
the system, the so-called ``Quantum Zeno effect.''~\cite{misra77,chiu77,sakurai94}
More recently it was predicted 
that an enhancement of decay due to frequent measurements could be observed 
under somewhat more general conditions, which was named the ``Anti-Zeno 
effect.''~\cite{schieve89,kofman96,kofman00,facchi01}
We report here the first observation of both the Zeno and Anti-Zeno 
effects by repeated measurements during the non-exponential period of an 
unstable quantum system.

Our experiment consists of ultra-cold sodium atoms in an accelerated standing 
wave of light which creates an optical potential of the form
$V_0 \cos[2 k_{\text{L}} x - k_{\text{L}} a t^2]$, 
where $V_0$ is the amplitude of the potential, $k_{\text{L}}$ is the wave number of the light forming 
the potential, $x$ is the position in the laboratory frame, $a$ is the acceleration and $t$ is time.  
Transformation to the frame accelerated with the potential $(x')$ yields the form 
$V_0 \cos [2 k_{\text{L}} x'] - m a x'$ in which a constant inertial force on the atom with mass $m$ is 
apparent.  Classically, for a given amplitude $V_0$ and small enough acceleration $a$ atoms 
can be trapped inside this tilted `washboard potential' and be accelerated along with it.  
Quantum mechanically, an atom which is classically bound can escape from this 
trapped state into the continuum via tunneling~\cite{niu96,wilkinson96,bharucha97}.  The system is therefore unstable, 
and the decay would be expected to follow the universal exponential decay law.  It was 
shown, however, that deviations from exponential behavior should occur for short 
times, owing to the initial reversibility of the decay process~\cite{winter61,fonda78,niu98}.
Our system is the only 
one in which these predicted deviations have been observed~\cite{wilkinson97}.  Improvements to our 
previous setup have now allowed us to explore parameter regimes for which much 
stronger deviations from the exponential behavior occur.  This has enabled us to study 
the effect of repeated measurements on the decay of the system~\cite{misra77,chiu77,sakurai94}.
Even though 
measurement-induced suppression of the dynamics of a two-state driven system has 
been observed~\cite{itano90,kwiat95}, no such effect was measured on an unstable system. Frequent 
measurements during the decay of an unstable system are predicted to reduce or 
enhance the decay rate, depending on the measurement interval.

The experimental setup resembles the one used previously to study deviations 
from exponential decay and will be described only briefly.  Several steps were 
necessary to prepare the initial condition.  We started by cooling and trapping 
approximately $3\cdot 10^5$ sodium atoms in a magneto-optical trap, followed by a stage of 
molasses cooling~\cite{cohen-tannoudji92}.  After this stage the distribution had a typical Gaussian width of 
$\sigma_{\text{x}} = 0.3$~mm in position and
$\sigma_{\text{v}} = 6 \, v_{\text{rec}}$ in velocity,
where $v_{\text{rec}} = 3$~cm/s is the single-photon
recoil velocity.  After switching off the cooling and trapping fields the 
interaction beams were turned on.  The interaction potential was a standing wave 
created by two linearly polarized counter-propagating laser beams with parallel 
polarization vectors.  The light was far detuned from the 
($3\text{S}_{1/2}) \leftrightarrow (3\text{P}_{3/2}$) transition in 
order to avoid electronic excitation and the resulting spontaneous emission.  Detunings 
typically ranged from 40 to 60~GHz and the power in each of the beams was adjusted up 
to 150~mW.  The beams were spatially filtered and focused to a beam waist of 1.8~mm 
at the position of the atomic cloud.  Due to the large initial momentum spread of the 
atomic cloud, switching on the interaction potential populated several of the lower 
energy bands.  Atoms projected into the lowest band are trapped within the potential 
wells whereas atoms in the second band are only partially trapped.  Atoms in even 
higher bands have energies well above the potential and hence are effectively free.  To 
empty all but the lowest band, the standing wave was then accelerated to a velocity of 
$v_0 = 35 \, v_{\text{rec}}$ by
linearly chirping the frequency of one of the counter-propagating beams 
while keeping the frequency of the other beam fixed.  As discussed in our previous 
work~\cite{niu96,wilkinson96,bharucha97}, accelerating the potential leads to a loss of population in the lower bands due 
to tunneling of atoms into higher untrapped bands.  The transport acceleration $a_{\text{trans}}$ was 
chosen to maximize tunneling out of the second band while minimizing losses from the 
first trapped band.  This ensured that after the initial acceleration only the first band still 
contained a significant number of atoms.  After reaching the velocity $v_0$ the acceleration 
was suddenly increased to a value $a_{\text{tunnel}}$ where appreciable tunneling out of the first 
band occurred.  This large acceleration was maintained for a period of time $t_{\text{tunnel}}$ after 
which time the frequency chirping continued again at the decreased rate corresponding 
to $a_{\text{trans}}$.  This separated in momentum space the atoms that were still trapped in the 
lowest band from those in higher bands.  After reaching a final velocity of $75 \, v_{\text{rec}}$ the 
interaction beams were switched off suddenly. A diagram of the velocity profile
versus time is shown in Fig.~1(a).

\begin{figure}
  \begin{center}
    \includegraphics[bb=20 260 570 535, width=8cm, keepaspectratio, clip]{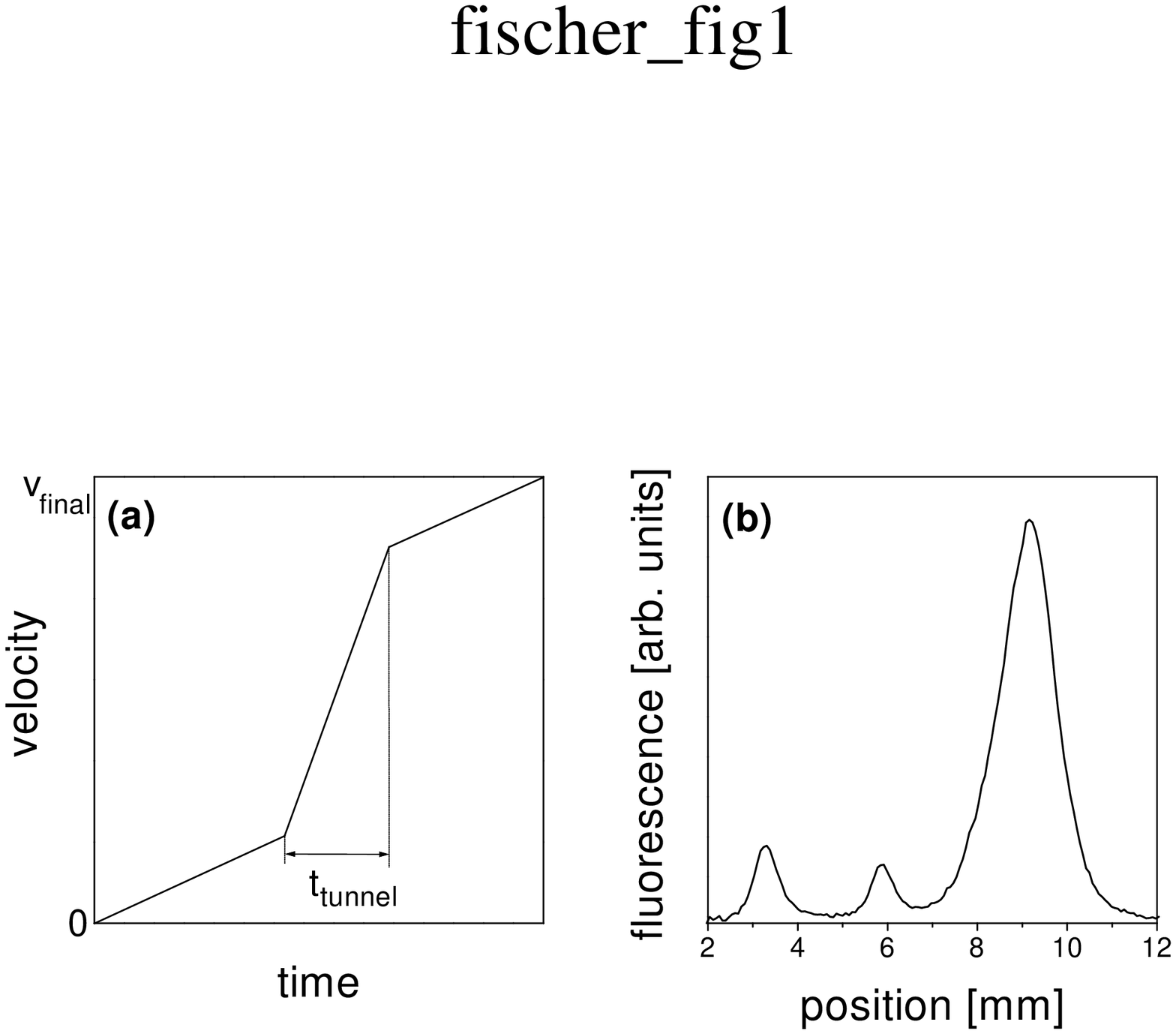}
  \end{center}
    \caption{
    Part (a) shows a diagram of the acceleration sequence.
    Part (b) shows a typical integrated spatial distribution of atoms after the time of ballistic expansion.
    The large peak on the right is the part of the atomic cloud that was not trapped during
    the initial acceleration. The center peak indicates the atoms that tunneled out of the
    optical potential during the fast acceleration period.
    The leftmost peak corresponds to atoms that remained trapped during the entire sequence.
    The survival probability is the area under the left peak normalized by the sum of the areas under
    the left and center peak.}
\end{figure}

In the detection phase we 
determined the number of atoms that were initially trapped and what fraction remained 
in the first band after the tunneling sequence.  After an atom tunneled out of the 
potential during the sequence, it would maintain the velocity that it had at the moment 
of tunneling.  Turning off the light beams allowed the atoms to expand freely. During 
this period of ballistic expansion each atom moved a distance proportional to its 
velocity. Due to the difference in final velocity, trapped and tunneled atoms separated 
and could be spatially resolved.  For imaging purposes the cooling beams were turned 
back on with no magnetic field gradient present.  This temporarily restricted movement 
of the atoms in a `freezing molasses', while the fluorescence was imaged onto a
charge-coupled-device camera.
Regions of the two-dimensional image were then integrated to 
obtain the desired fraction of remaining atoms over the number of initially trapped 
atoms. A typical integrated distribution is shown in Fig.~1(b). For this trace, about one
third of the initially trapped atoms have tunneled out
of the well during the fast acceleration duration.  

\begin{figure}
  \begin{center}
    \includegraphics[bb=20 260 570 535, width=8cm, keepaspectratio, clip]{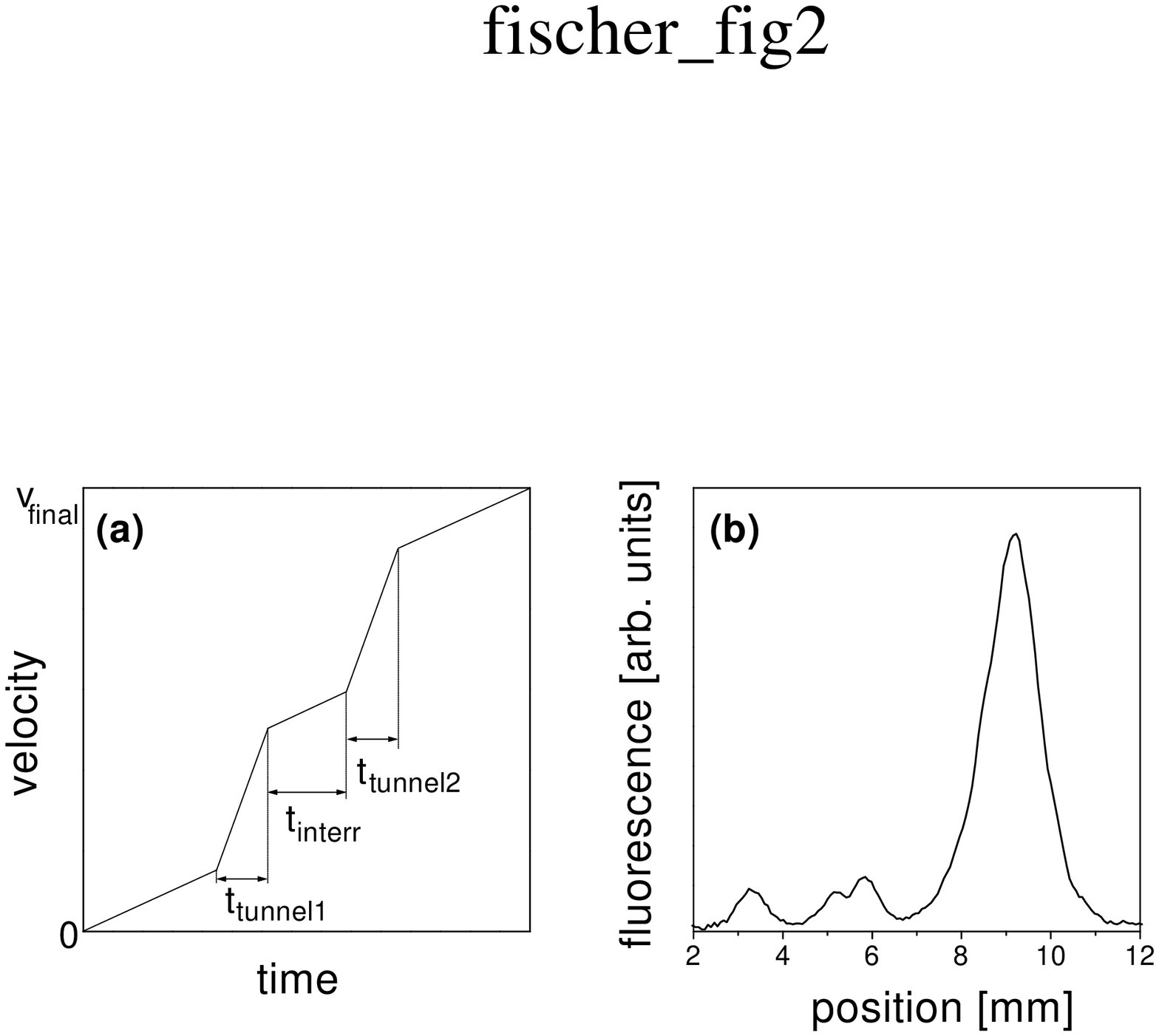}
  \end{center}
    \caption{
    Part (a) shows a diagram of the interrupted acceleration sequence. The total tunneling time
    is the sum of all the tunneling segments.
    Part (b) shows a typical integrated spatial distribution of atoms after the time of ballistic expansion.
    The peaks can be identified as in Fig.~1. However, the area containing the tunneled fraction of the atoms
    is now composed of two peaks. 
    Atoms that left the well during the first tunneling segment are offset in velocity
    from the ones having left during the second period of tunneling. The amount of
    separation is equal to the velocity increase of the well during the interruption segment.}
\end{figure}

We observed the decay of the unstable system by repeating the experiment for 
various tunneling durations $t_{\text{tunnel}}$, holding the other parameters of the sequence fixed.  
The focus of this work, however, was the effect of measurements on the system decay 
rate.  The quantity to be measured was the number of atoms remaining trapped in the 
potential during the tunneling segment.  This measurement could be realized by 
suddenly interrupting the tunneling duration by a period of reduced acceleration
$a_{\text{interr}}$, 
as indicated in Fig.~2(a).  During this interruption tunneling was negligible and the atoms 
were therefore transported to a higher velocity without being lost out of the well. This 
separation in velocity space enabled us to distinguish the remaining atoms from the ones 
having tunneled out up to the point of interruption, as can be seen in Fig.~2(b).
By switching the acceleration back 
to $a_{\text{tunnel}}$ the system was then returned to its unstable state.  The measurement of the 
number of atoms defined a new initial state with the remaining number of atoms as the 
initial condition.  The system must therefore start the evolution again with the same 
non-exponential decay features.  The requirements for this interruption section were 
very similar to those during the transport section, namely the largest possible 
acceleration while maintaining negligible losses for atoms in the first band. Hence 
$a_{\text{interr}}$ was chosen to be the same as $a_{\text{trans}}$.

\renewcommand{\textfraction}{.01}
\renewcommand{\topfraction}{.99}
\renewcommand{\dbltopfraction}{.99}
\renewcommand{\floatpagefraction}{.99}
\renewcommand{\dblfloatpagefraction}{.99}
\begin{figure}
  \begin{center}
    \includegraphics[bb=75 190 490 560, width=8cm, keepaspectratio, clip]{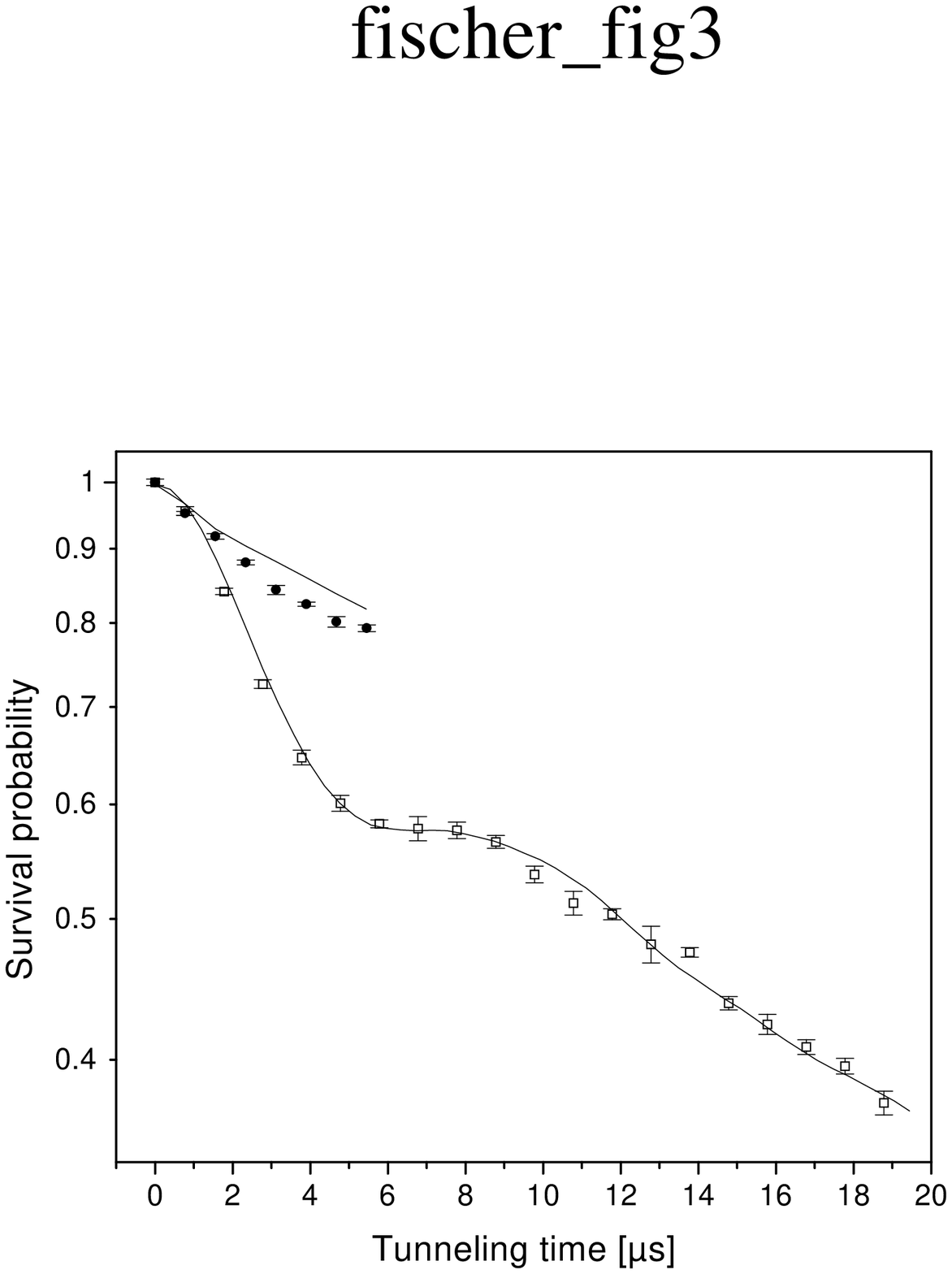}
  \end{center}
    \caption{
    Probability of survival in the accelerated potential as a function of 
    duration of the tunneling acceleration.  The hollow squares show the
    non-interrupted sequence, the solid circles show the sequence with interruptions of 
    50~$\mu$s duration every 1~$\mu$s.  The error bars denote the error of the mean.  The 
    data have been normalized to unity at $t_{\text{tunnel}} = 0$ in order to compare to the 
    simulations.  The solid lines are quantum mechanical simulations of the 
    experimental sequence with no adjustable parameters.  For these data the 
    parameters were: $a_{\text{tunnel}} = 15,000 \, \text{m}/\text{s}^2$, 
    $a_{\text{interr}} = 2,000 \, \text{m}/\text{s}^2$, $t_{\text{interr}} = 50 \, \mu$s and 
    $V_0/h = 91$~kHz, where $h$ is Planck's constant.}
\end{figure}

Figure~3 shows the dramatic effect of frequent measurements on the decay 
behavior.  The hollow squares indicate the decay curve without interruption.  As 
pointed out by Misra and Sudarshan~\cite{misra77}, one can take advantage of the slow initial decay 
in order to inhibit the decay altogether by frequently measuring the system at very short 
time intervals.  They named this suppression of decay the Quantum Zeno effect.  The 
solid circles in Fig.~3 depict the measurement of the survival probability in which after 
each tunneling segment of 1~$\mu$s an interruption of 50~$\mu$s duration was inserted.  Only 
the short tunneling segments contribute to the total tunneling time. The survival 
probability clearly shows a much slower decay than the corresponding system measured 
without interruption.  Care was taken to include the limited time response of the 
experimental setup into the analysis of the data.  The response time was limited by 
electronic and electro-optic devices used in the experiment.  The frequency response 
was measured and the resulting transfer function was used to calibrate the response of 
the optical potential to a desired change in acceleration.  This ensured that only sections 
were included for which tunneling was substantial and established a lower bound for 
the actual tunneling duration. This effect was taken into account for the curves in 
Fig.~3.  Also indicated as solid lines are quantum mechanical simulations of the decay 
by numerically integrating Schr\"odinger's equation for the experimental sequence and 
determining the survival probability numerically.  The simulations contained no 
adjustable parameters and are in good agreement with the experimental data. We 
attribute the seemingly larger decay rate for the Zeno experiment as compared to the 
simulation to the under-estimate of the actual tunneling time.  This suggests that in 
reality the decay might be even slower than indicated by the experimental data points.

\begin{figure}
  \begin{center}
    \includegraphics[bb=70 190 490 560, width=8cm, keepaspectratio, clip]{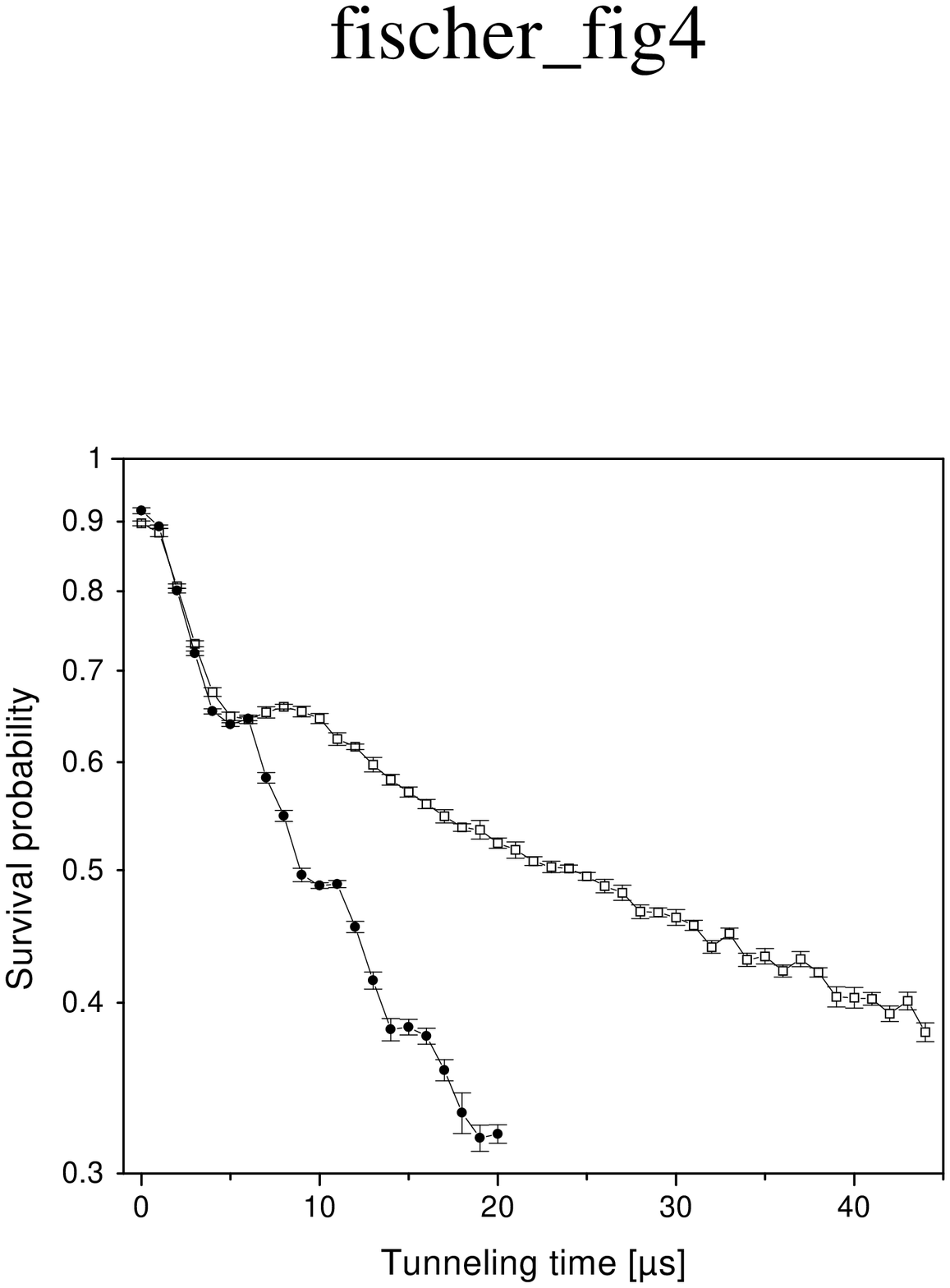}
  \end{center}
    \caption{
    Survival probability as a function of duration of the tunneling 
    acceleration.  The hollow squares show the non-interrupted sequence, the solid 
    circles show the sequence with interruptions of 40~$\mu$s duration every 5~$\mu$s.  The 
    error bars denote the error of the mean.  The experimental data points have 
    been connected by solid lines for clarity.  For these data the 
    parameters were: $a_{\text{tunnel}} = 15,000 \, \text{m}/\text{s}^2$, 
    $a_{\text{interr}} = 2,800 \, \text{m}/\text{s}^2$, $t_{\text{interr}} = 40 \, \mu$s and 
    $V_0/h = 116$~kHz.}
\end{figure}

The shape of the uninterrupted decay curve suggests yet another option for 
changing the decay behavior of an unstable system.  After an initial period of slow 
decay the curve shows a steep drop as part of an oscillatory feature, which for longer 
time damps away to show the well-known exponential decay.  If the system was to be 
interrupted right after the steep drop, one would expect an overall decay that is faster 
than the uninterrupted decay~\cite{kofman00}.  In contrast to the slower decay for the Zeno effect this 
prediction was named the Anti-Zeno effect.  The solid circles in Fig.~4 show such a 
decay sequence, where after every 5~$\mu$s of tunneling the decay was interrupted by a 
slow acceleration segment.  As in the Zeno-case, these interruption segments force the 
system to repeat the initial non-exponential decay behavior after every measurement.  
Here, however, the tunneling segments between the measurements are chosen longer in 
order to include the periods exhibiting fast decay.  The overall decay is much faster than 
for the uninterrupted case, indicated by the hollow squares in the same figure.

\begin{figure}
  \begin{center}
    \includegraphics[bb=70 190 490 560, width=8cm, keepaspectratio, clip]{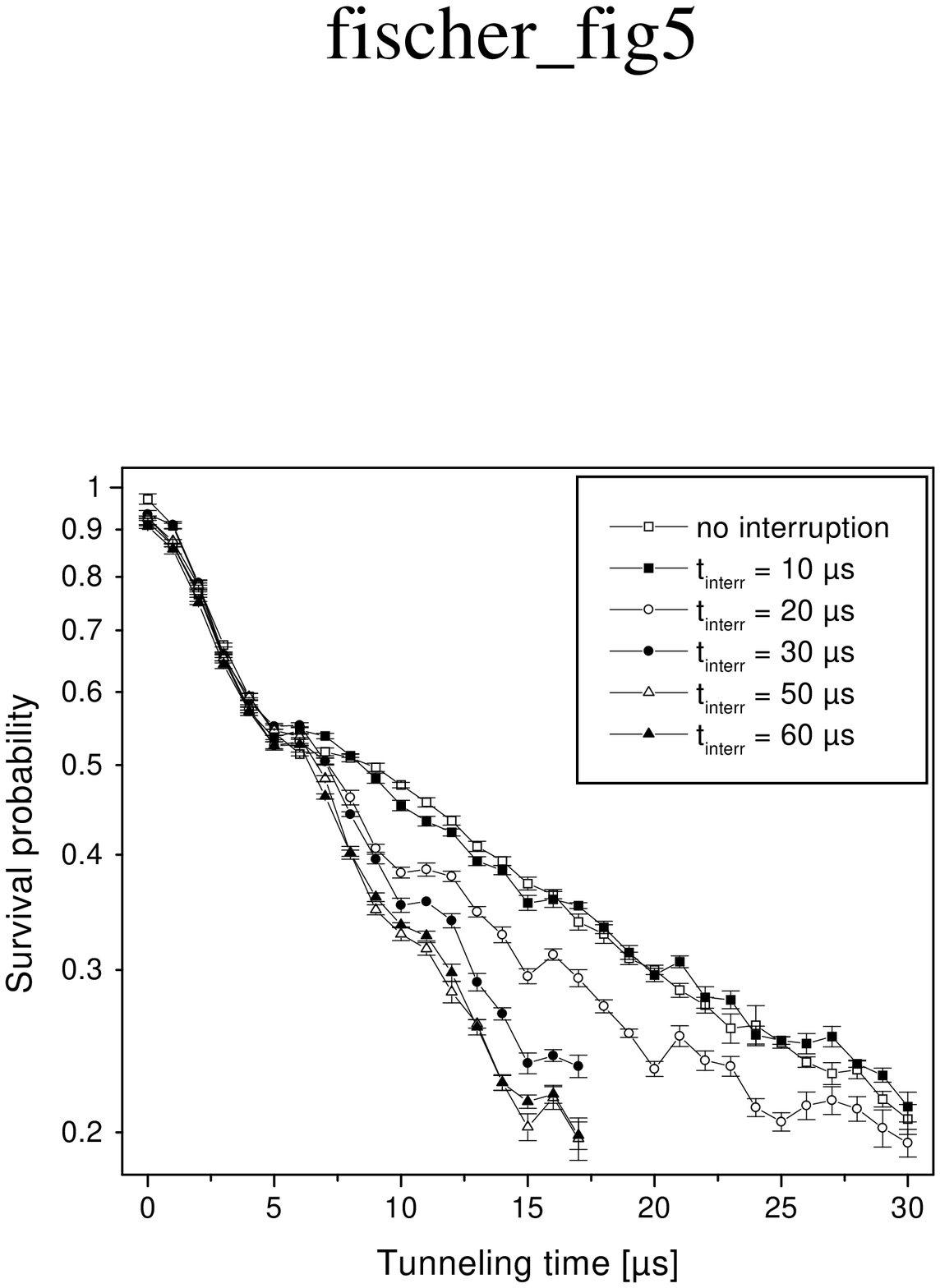}
  \end{center}
    \caption{
    Survival probability as a function of duration of the tunneling 
    acceleration.  The hollow squares show the non-interrupted sequence, other 
    symbols indicate the sequence with a finite interruption duration after every 5~$\mu$s 
    of tunneling.  The error bars denote the error of the mean.  A further increase of 
    the interruption duration than as indicated does not result in a further change of 
    the decay behavior.  The experimental data points have been connected by 
    solid lines for clarity.  For these data the 
    parameters were: $a_{\text{tunnel}} = 15,000 \, \text{m}/\text{s}^2$, 
    $a_{\text{interr}} = 2,000 \, \text{m}/\text{s}^2$ and $V_0/h = 91$~kHz.}
\end{figure}

The key to observing the Zeno and Anti-Zeno effects is the ability to measure 
the state of the system in order to repeatedly redefine a new initial state.  In our case the 
measurement is done by separating in momentum space the atoms still left in the 
unstable state from the ones that decayed into the reservoir. In order to distinguish the 
two classes of atoms, they must have a separation of at least the size of the momentum 
distribution of the unstable state, which in our case is the width of the first Brillouin 
zone of $\delta p = 2 m v_{\text{rec}}$.
The time it takes for an atom to be accelerated in velocity by this 
amount is the Bloch period $\tau_b = 2 v_{\text{rec}} / a_{\text{interr}}$,
assuming an acceleration of $a_{\text{interr}}$.  An 
interruption shorter than this time will not resolve the tunneled atoms from those still 
trapped in the potential and therefore results in an incomplete measurement of the atom 
number.  To investigate the effect of the interruption duration we repeated a sequence to 
measure the Anti-Zeno effect for varying interruption durations while holding all other 
parameters constant.  Fig.~5 displays the results of this measurement, interrupting the 
decay every 5~$\mu$s with an acceleration of $a_{\text{interr}}$ of 
$2000 \, \text{m}/\text{s}^2$.  The hollow squares show 
the uninterrupted decay sequence as a reference.  For an interruption duration smaller 
than the Bloch period of 30~$\mu$s the measurement of the atom number is incomplete and 
has little or no effect.  For a duration longer than the Bloch period the effect saturates 
and results in a complete restart of the decay behavior after every interruption. Even 
though this method of interruption is not an instantaneous measurement of the state of 
the unstable system, we can still accomplish the task of redefining the initial state by 
first switching the system from an unstable to a stable one, then in a finite time perform 
the measurement and finally switching the system back to be unstable again.

One of us (M.G. R.) thanks Gershon Kurizki and Abraham Kofman for helpful discussions. This work 
was supported by NSF, the R. A. Welch Foundation and the Sid W. Richardson Foundation.

\end{document}